\documentclass[a4paper,11pt]{article}
\pdfoutput=1
\usepackage{jheppub} 
\usepackage{graphicx,color,rotating}
\usepackage{hyperref}
\usepackage{epsfig,color}
\usepackage{slashed}
\usepackage{amsfonts}
\usepackage{ulem}

\usepackage{tabularx}
\usepackage{graphicx}
\usepackage{adjustbox}
\usepackage{tabularx}

\begin{document}
\title{Type Ia Supernovae induced by Primordial Black Hole via Dark First-Order Phase Transition}

\renewcommand{\thefootnote}{\arabic{footnote}}

\author{
Pin-Jung Chen$^{1}$ and 
Po-Yan Tseng$^{1}$}
\affiliation{
$^1$ Department of Physics, National Tsing Hua University,
101 Kuang-Fu Rd., Hsinchu 300044, Taiwan R.O.C. \\
}
\date{\today}

\abstract{
Primordial black hole (PBH) with mass $10^{-15}\leq M_{\rm PBH}/M_{\odot}\leq 10^{-10}$ is currently beyond the sensitives via neither microlensing nor black hole (BH) evaporation.
A novel scenario
when a PBH, $10^{-14}\leq M_{\rm PBH}/M_{\odot}\leq 10^{-11}$,
transits through a white dwarf (WD) made up of carbon and oxygen, Bondi-Hoyle-Lyttleton (BHL) accretion in a reactive medium creates a shock wave, which generates direct detonation ignition in the WD core and then leads to thermonuclear supernovae (SNe Ia), has been proposed.
The aim of this study was to impose constraints on the PBH fraction $f_{\rm PBH}$ of dark matter (DM) via comparing the SN Ia event rates between PBH hypotheses and observational data.
For PBH fraction less than unity, we found
the observed event rate prefers PBH mass regions in 
$7.6\times 10^{-13}\leq M_{\rm PBH}/M_{\odot}\leq 6.1\times 10^{-12}$
under the Navarro–Frenk–White (NFW) profile.
Meanwhile, the aforementioned PBH mass and abundance can be efficiently generated via the production mechanism of cosmological first-order phase transition (FOPT) in dark sector associates with $\mathcal{O}({\rm MeV})$ energy scale, giving rise complementary signal of stochastic gravitational waves (GWs) with $10^{-6}$ Hz to $10^{-5}$ Hz peak frequencies, which can be probed by future $\mu$Ares GW interferometer.
}
\maketitle

\bigskip
\section{Introduction}
PBH is a hypothetical celestial object, have formed before the star and galaxy formations~\cite{Zeldovich:1967lct,Hawking:1971ei},
serves as a dark matter (DM) candidate~\cite{Hawking:1971ei,Chapline:1975ojl,Khlopov:2008qy,Carr:2016drx,Carr:2020gox,Carr:2020xqk,Green:2020jor,Bertone:2004pz}. 
%
From the observational perspective, GW signatures generated by merging of BH binary, was firstly detected and confirmed by LIGO and VIRGO~\cite{LIGOScientific:2016aoc}, so that there has been a renewed interest in PBHs.
Because gravitational waves could propagate cosmological distance without attenuation by interstellar mediators~\cite{Sasaki:2018dmp}, they are utilized as a new method to explore beyond recombination epoch, and thus growing investigations in the PBH formation paradigms in the early Universe through various mechanisms: the collapse of overdensity coming from the primordial fluctuations after inflation~\cite{Carr:1974nx,Sasaki:2018dmp}; a cosmological first-order phase transition (FOPT) may accumulate sufficient energy density within Schwarzschild radius via bubble wall collision~\cite{Hawking:1982ga,Kodama:1982sf,Moss:1994iq,Konoplich:1999qq}; FOPT incorporates dark sector particles, in which Fermi Balls (FBs) behave as progenitor stars collapsing to form PBHs~\cite{Baker:2021nyl,Gross:2021qgx,Kawana:2021tde,Marfatia:2021hcp}.

Current literature on PBH pays particular attention to the constraints on PBH abundance as the fraction, $f_{\rm PBH}$, to total DM relic, depending on the mass range and properties of PBH.
When a PBH mass is less than about $10^{-16}M_\odot$~\cite{Carr:2020gox}, which can be probed by the Hawking radiation emanating from an evaporating PBH~\cite{Marfatia:2021hcp} and contribution to the cosmic gamma-ray or neutrino flux; on the other hand, when a PBH has mass greater than about $10^{-11}M_\odot$~\cite{Carr:2020gox}, the most stringent bound comes from microlensing effect, where gravity lensing enhance the luminosity of electromagnetic source emitted from background stars due to PBH transiting through the line-of-sight~\cite{Croon:2020ouk}. However, the mass window, $10^{-16}\lesssim M_{\rm PBH}/M_\odot\lesssim 10^{-11}$ remains intact, hampering from the amount of Hawking radiation is inversely proportional to the mass of PBH, and the occurrence of microlensing requires the Schwarzschield radius of PBH to be greater than the optical wavelength, so that current observations have no sensitivities to the aforementioned mass window. Whereas based on future observation, people proposed that using the femtolensing of Gamma-ray Bursts, with two detectors separation in two astronomical units, has potential to cover this PBH mass window~\cite{Jung:2019fcs}.

Alternatively, 
this study aims to fill the gap
by considering a asteroid-mass PBH pass through a white dwarf (WD), which creates a shock wave via the supersonic flow toward the gravitational center (Bondi-Hoyle-Lyttleton (BHL) accretion), such that it initiates the normal thermonuclear supernovae (SN Ia), and turn out the current Type Ia supernovae data is able to constrain the fraction of asteroid PBH abundance.
The detonation ignition criteria can be satisfied when the closest distance between a free falling PBH and a WD less than $R_m$~\cite{Steigerwald:2021vgi}, which specifically depends on masses of PBH and WD.
Eventually, the SN Ia event rate induced by PBHs, inside the galaxy is estimated by taking the ignition cross section $\pi R^2_m$ and
WD distribution in the entire galaxy, which utilized the extrapolation of 
observed WDs mass distribution within 100 pc of our solar system~\cite{Kilic_2020} and assumed the number density of WDs is proportional to the number density of stellar.
We also choose the density distribution of PBHs following the NFW, isothermal, Kra, and Moore profiles to demonstrate the variance of the event rate.

Currently, there is no consensus referring to the mechanism of PBH formation.
In this work, we discuss plausible production scenario of asteroid-mass PBHs in the early universe from FOPT in dark sector, associating with dark scalar $\phi$ and dark fermion $\chi$~\cite{Hong:2020est}. 
The finite temperature potential of $\phi$ induces the FOPT, and result in the Universe transferring from false vacuum into true vacuum, undergo the bubble nucleation and expansion. 
On top of that, $\phi$ developed non-zero vacuum expectation value and couples to $\chi$, making $\chi$ gain additional mass in the true vacuum.
Suppose the characteristic temperature of FOPT is less than the $\chi$ mass difference, 
the $\chi$'s located in false vacuum do not carry sufficient kinetic energy to enter the true vacuum, result in trapping $\chi$'s
in the false vacuum.
%
These trapped $\chi$'s are aggregated due to the shrinking volume of false vacuum, eventually they form the Fermi balls (FBs)~\cite{Hong:2020est}, in which Fermi energy dominates and balance with the vacuum pressure.
Afterward, the attractive Yukawa potential, whose magnitude increases as FB temperature decrease, surpasses the Fermi energy and makes FBs collapse into PBHs~\cite{Kawana:2021tde}, where the uncertainty of FB cooling rate relies on the model detail that has been discussed in Ref.~\cite{Kawana:2022lba,Lu:2022jnp}.
We found the corresponding temperature of FOPT around $\mathcal{O}({\rm MeV})$ can generate the PBH mass between $10^{-13}M_\odot$ to $10^{-11}M_\odot$ and right amount of fraction abundance to be consistent with current SN Ia event rate, meanwhile FOPT produces gravitational waves peaking around $10^{-5}$ Hz and falling into the sensitivity of future $\mu$Ares~\cite{Sesana:2019vho}.

This paper proceeds in the following way. We first content ourselves with a novel mechanism of the formation of PBHs by first order phase transition in section~\ref{sec:FOPT}. Then the event rate of SN Ia triggered by PBHs is calculated in section~\ref{sec:PBH_SN}. Finally, the constraints on PBHs in accordance with the observed SN Ia event rate are discussed and summary in section~\ref{sec:even_rate} and \ref{sec:summary}, respectively.

\bigskip

\section{First-order phase transition and PBH formation}
\label{sec:FOPT}



We consider a novel PBH production mechanism, in which Dirac fermion particles in dark sector form Fermi-balls (FBs) in advance, as an intermediate state due to a FOPT in the early Universe~\cite{Hong:2020est}, afterward FB collapses into PBH attributing to the attractive Yukawa force from a dark scalar~\cite{Kawana:2021tde}. One model realizing this scenario is depicted below.

Let $\phi$ be a real scalar field 
and $\chi$ be a Dirac fermion particles in the dark sector. Then the simplest Lagrangian which can realise the aforementioned mechanisms is~\cite{Kawana:2021tde,Huang:2022him}
\begin{eqnarray}
\label{eq:lagrangian}
\mathcal{L} = -\frac{1}{2}\partial_{\mu}\phi\partial^{\mu}\phi - V_{\rm eff}(\phi,T) + \bar{\chi}(i\slashed{\partial}-m)\chi - g_{\chi}\phi\bar{\chi}\chi.
\end{eqnarray}
It will be manifest later that $m$, the bare mass of $\chi$, is added to the Lagrangian with the purpose of satisfying the first stability condition for FB in Eq.~(\ref{eq:stability}).
Take $V_{\rm eff}$ dictating the cosmological FOPT, to be the quartic effective thermal potential
at temperature $T$ in the dark sector~\cite{Dine:1992wr, Adams:1993zs}
\begin{eqnarray}
\label{eq:potential}
V_{\rm eff}(\phi,T) = D(T^2-T_0^2)\phi^2 - (AT+C)\phi^3 + \frac{\lambda}{4}\phi^4\,,
\end{eqnarray}
here $2D(T^2-T^2_0)$ representss the square of thermal mass of $\phi$.
When $T$ falls below the critical temperature $T_c$, the two local minima become nondegenerate, and the vacuum transferring from the local minimum at $\langle\phi\rangle = 0$ (called \textit{false vacuum}) to the global minimum at $\langle\phi\rangle = v_{\phi}(T)$ (called \textit{true vacuum}) is triggered. 
The characteristic energy scale of FOPT is defined as zero-temperature potential energy at the global minimum, $B=-V_{\rm eff}(v_\phi,0)$~\cite{Marfatia:2021twj}.
After trading $T_0$ by $B$, we are able to select input parameters: $\lambda$, $A$, $B$, $C$, and $D$, to determine the evolution of FOPT.
We further define the percolation temperature of FOPT $T_\star$, associating with $1/e \simeq 0.37$ of space volume remain occupied by false vacuum~\cite{Wang:2020jrd}, and it also corresponds to the formation temperature of FBs.


The last two terms of Eq. (\ref{eq:lagrangian}) implies that $\chi$ has masses $|m|$ and $|m+g_\chi v_\phi|$ in the false and true vacua, respectively~\cite{Huang:2022him}. If the mass gap is much larger than the order of average thermal kinetic energy of $\chi$’s
\begin{eqnarray}
\label{eq:mass_gap}
g_\chi v_\phi \gg T_\star,
\end{eqnarray}
most of $\chi$’s/$\bar{\chi}$'s do not have sufficient thermal kinetic energy to penetrate from the false vacuum into the true vacuum,
result in, trapping them in the false vacuum remnants~\cite{Kawana:2021tde}. $F^{\rm trap}_\chi$ denotes the fraction of $\chi$/$\bar{\chi}$ remains in the false vacuum~\cite{Hong:2020est,Huang:2022him}.
The assumption of nonzero asymmetry $\eta_\chi \equiv (n_\chi-n_{\bar{\chi}})/s$ in number density between $\chi$ and $\bar{\chi}$ (normalized to the entropy density $s$) is implemented to prevent $\chi \bar{\chi} \to \phi \phi$ complete annihilation, and this can be preserved by assigning a $U(1)$ symmetry. Explicit mechanism producing $\chi$-asymmetry have been proposed in Ref.~\cite{Hong:2020est,Huang:2022him}.
Afterward,
the true vacuum bubbles stop nucleating among the false vacuum and hence the false vacuum stops splitting, occurring when the later has shrunk to the critical volume $V_{\rm crit} = \frac{4\pi}{3}R(T_\star)^3$, 
making the bubble nucleation probability less than $\mathcal{O}(1)$ within this volume~\cite{Kawana:2021tde}.
Base on the fact that each critical volume evolves into a FB , the number density of FBs and total number of dark fermions inside a FB are given by\begin{eqnarray}
\label{eq:number_density}
n_{\rm FB}(T_\star) = \frac{0.37}{V_{\rm crit}},~~~~
Q_{\rm FB}=F^{\rm trap}_\chi \frac{\eta_\chi s(T_\star)}{n_{\rm FB}(T_\star)}\,,
\end{eqnarray}
respectively. Here we simply assume trapping efficiency equals to unity.

The total energy $E_{\rm FB}$ of a FB is the summation of: the Fermi-gas degeneracy pressure with thermal excitation; the surface tension; the bulk energy; and the Yukawa energy potential energy~\cite{Hong:2020est,Kawana:2021tde}\,.
In the absence of Yukawa interaction, one can find the minimum of $E_{\rm FB}$ by adjusting the radius of FB, which yields the stable solution and determines the radius and mass of FB.
The leading order approximation gives~\cite{Marfatia:2021hcp}
$$
M_{\rm FB}\approx Q_{\rm FB}\left( 12 \pi^2 V_0 \right)^{1/4}\,,~~R_{\rm FB}\approx Q^{1/3}_{\rm FB}\left[ \frac{3}{16}\left( \frac{3}{2\pi}  \right)^{2/3} \frac{1}{V_0} \right]^{1/4}\,.
$$
with $V_0 \equiv V_{\rm eff}(0,T_\star)-V_{\rm eff}(v_\phi(T_\star),T_\star)$.
Further stability conditions for FBs are~\cite{Marfatia:2021hcp}
\begin{eqnarray}
\label{eq:stability}
\begin{split}
\frac{dM_{\rm FB}}{dQ_{FB}} < m + g_\chi v_\phi\,,~~
\frac{d^2M_{\rm FB}}{d^2Q_{\rm FB}} < 0.
\end{split}
\end{eqnarray}
The Yukawa potential contributes negatively into the $E_{\rm FB}$ and is mediated by $\phi$, which obtains a thermal mass $M^2_\phi = \frac{d^2V_{\rm eff}}{d\phi^2}|_{\phi = 0} = 2D(T^2-T_0^2)$~\cite{Marfatia:2021hcp}.
At $T_\star$, Yukawa interaction length $L_\phi \equiv M^{-1}_\phi$ is shorter than the mean separation of $\chi$s in FB, so ignoring the Yukawa energy is validated.
Yukawa potential remains subdominant until the temperature drops to $T_\phi$, when the $L_\phi$ is comparable to the mean separation of $\chi$'s~\cite{Kawana:2021tde}
\begin{eqnarray}
\label{eq:mean_separation}
L_\phi \approx \left( \frac{\bar R_{\rm FB}^3}{Q_{\rm FB}} \right)^{1/3},
\end{eqnarray}
the Yukawa potential energy prevails over the $E_{\rm FB}$ and the negative sign leads to the collapse of FBs into PBHs. 
Therefore, the mass $M_{\rm PBH}$ of a PBH derived by these mechanisms is $M_{\rm PBH} = M_{\rm FB}({T_\phi})$ and the number density $n_{\rm PBH}=n_{\rm FB}|_{T_\star}s(T_\phi)/s(T_\star) $ according to the adiabatic evolution of the Universe~\cite{Kawana:2021tde}.
At present Universe with temperature $T_0=0.235~{\rm meV}$, the PBH number and relic density are given by
\begin{eqnarray}
\left.\left( \frac{n_{\rm PBH}}{s} \right) \right |_{T_0}
=\left.\left( \frac{n_{\rm PBH}}{s} \right) \right |_{T_\phi}\,,~~ \Omega_{\rm PBH}h^2 
= \frac{(M_{\rm PBH}\, n_{\rm PBH})|_{T_0}}{3M^2_{\rm Pl}(H_0/h)^2}\,, 
\end{eqnarray}
where $H_0$ is the Hubble constant.
In this work, we focus on $M_{\rm PBH}/M_\odot > \mathcal{O}(10^{-12})$, and thus the PBH mass change from $T_\phi$ to $T_0$ due to Hawking radiation is insignificant. 



Caveat regarding the $T_\phi$ and FB cooling rate, depend highly on details of the dark sector model and how they interact with standard model (SM) sector.
For instant, the FBs cool down through portal couplings, mixing between $\phi$ and SM Higgs, and then by emitting light SM particles, such as electrons and neutrinos~\cite{Kawana:2021tde, Hong:2020est}. 
Suppose the mixing is sufficiently large, the FB cooling period is much smaller than the Hubble time scale, the FBs can follow the temperature of the Universe till today~\cite{Witten:1984rs}.
However, alternative scenario, late-forming PBH after the CMB epoch, is also plausible, where the cooling only relies on the evaporation for dark sector particles, but this circumvents the constraint on PBH from CMB accretion bound~\cite{Lu:2022jnp}. FB cooling mechanism is beyond the scope of this work, and we will investigate it in future studies.



\bigskip

\section{PBHs trigger Type Ia SN}
\label{sec:PBH_SN}

\bigskip

\subsection{Thermonuclear detonation ignition}

We follow the detonation ignition in a reactive medium for Type Ia supernovae explosion~\cite{Steigerwald:2021vgi},
which can be achieved when a PBH passes through the reactive medium of a WD,
where supersonic flow onto the PBH gravitational center creates the shock wave,
known as Bondi-Hoyle-Lyttleton (BHL) accretion.
In general, the maximum radius inside WD associates with detonation ignition $R_m(M_{\rm WD},M_{\rm PBH})$ can be derived~\cite{Steigerwald:2021vgi}. 
As a result, considering a PBH carries the velocity $v_{\rm gal}$ in the galaxy falling into a WD from infinity, the effective ignition cross section is given by 
\begin{eqnarray}
    \pi R^2_m\left(\frac{v_{*{\rm esc}}(R_m)}{v_{\rm gal} }\right)^2\,,
\end{eqnarray}
including the focus effect from WD gravity.

The $R_m$ were determined from three criteria for inducing successful shock wave and self-sustained ignition detonation in the BHL accretion flow.
Base on the Zeldovich-von Neuman-Doring (ZND) model~\cite{ZND_vCJ,ZND}, the leading shock is followed by a reaction zone, which is separated by the induction length $\xi$. 
The bow-shaped BHL shock can be parameterized by the hyperbola function.
We can find a critical point $C$ on the shock, where the WD medium crossing the shock inside $C$ accretes into the BH.
If the pressure wave induced by the heat from the reaction zone can reach the shock and power it, self-sustained detonation can be realized. 
The sufficient condition is to check the induction length at the crucial point $C$, denoted as $\xi_c$.
Finally, the triple point is defined as the orthogonal projection from the reaction zone on the shock front.
Three criteria for self-sustained detonation: i) the postshock Mach number at the triple point form the end of the critical induction length need to be sonic, ii) the preshock velocity at this triple point has to exceed Chapman-Jourgeut velocity~\cite{ZND_vCJ,v_CJ}, iii) the $\xi_c$ should be at most of the critical impact parameter $b_c$, to maintain the reaction during the postshock state. For the last criterion, 
$\bar{\alpha}$ is a parameter of order one whose multiplication with the critical impact parameter serves as an upper bound for the critical induction length~\cite{Steigerwald:2021vgi}
\begin{eqnarray}
\label{eq:alpha}
\xi_{\perp c} < \bar{\alpha}b_c,
\end{eqnarray}
in this work, $\bar{\alpha} = 1$ is taken to be conservative and $\bar{\alpha}=10$ is taken to be more optimistic for the following analysis.

\bigskip

\subsection{Event rate}

We assume the PBH distribution follows the DM profiles and take NFW, Isothermal, Kra and Moore density distributions~\cite{Bertone:2004pz}
\begin{eqnarray}
\label{eq:DM density profile}
\rho_{\rm DM}(r_{\rm MW})= 
\frac{\rho_0}
{\left(\frac{r_{\rm MW}}{R_0} \right)^\gamma \left[1+\left( \frac{r_{\rm MW}}{R_0} \right)^\alpha \right]^{(\beta-\gamma)/\alpha} }\,,
\end{eqnarray}
with $(\alpha,\beta,\gamma,R_0)$ listed in Table~\ref{tab:DM_profile}.
Where $\rho_{\rm DM}$ is the density of DM halo, $r_{\rm MW}$ is the radius of a position in the Milky Way in spherical coordinates whose origin is the galactic centre, and $R_0$ is the normalisation scale radius.
Then the values of $\rho_0$
adjust the local DM density to be $\rho_{\rm DM}|_{r_{\rm MW}=8.25~{\rm kpc}}=0.4~{\rm GeV/cm^3}$.

\begin{table}[tb!]
  \centering

  \caption{Parameters for DM profiles. 
    \label{tab:DM_profile} }
  \medskip
  \begin{tabular}{l|lllll}
    \hline
        \hline
    & $\alpha$ &  $\beta$ & $\gamma$ & $R_0$ [kpc] & $\rho_0$ [${\rm GeV}/{\rm cm^3}$]\\
    \hline
NFW & 1.0 & 3.0 & 1.0 & 20.0 & 0.33 \\
Isothermal & 2.0 & 2.0 & 0.0 & 3.5 & 2.62 \\
Kra & 2.0 & 3.0 & 0.4 & 10.0 & 0.73 \\
Moore & 1.5 & 3.0 & 1.5 & 28.0 & 0.07 \\
    \hline \hline
  \end{tabular}
\end{table}

We further assume that the WD number density in the Milky Way is proportional to the mass density distribution of stars which distribute within the galactic disk,
$$
\frac{dn_{\rm WD}}{dM_{\rm WD}}(R_{\rm MW},Z_{\rm MW}) \propto \rho_{\rm stars}(R_{\rm MW},Z_{\rm MW})\,,
$$
where $R_{\rm MW}$ and $Z_{\rm MW}$ are the radius and height of a position in the Milky Way in the cylindrical coordinates whose origin is the centre of the galactic disk, relating to spherical radius as $r_{\rm MW}= \sqrt{R^2_{\rm MW}+Z^2_{\rm MW}}$. According to the observations,
the stellar density including only the thin disk contribution is parameterized by~\cite{deJong:2009iq}
$$
\rho_{\rm stars}(R_{\rm MW},Z_{\rm MW})\simeq \frac{0.038\times 10^9 \, M_\odot}{\rm kpc^3}
\times \left( e^{R_1/L_1} \exp\left( -\frac{R_{\rm MW}}{L_1}-\frac{|Z_{\rm MW}|}{H_1} \right) \right)\,,
$$
where the radius $R_1$ from the solar system to the galactic centre, length $L_1$ and height $H_1$ of the galactic disk have best-fit values $R_1=8.25~{\rm kpc}$, $L_1=2.6~{\rm kpc}$, and $H_1=0.25~{\rm kpc}$, respectively, which implies
\begin{eqnarray}
\frac{dn_{\rm WD}}{dM_{\rm WD}}(R_{\rm MW},Z_{\rm MW})= 
\left. \frac{dn_{\rm WD}}{dM_{\rm WD}}\right\vert_{\odot}
\times
\left( e^{R_1/L_1} \exp\left( -\frac{R_{\rm MW}}{L_1}-\frac{|Z_{\rm MW}|}{H_1} \right) \right)\,, \nonumber \\
\end{eqnarray}
where 
$$
\left. \frac{dn_{\rm WD}}{dM_{\rm WD}}\right\vert_{\odot} \equiv\frac{1}{({\rm 0.1\, kpc})^3}\,\left. \frac{dN_{\rm DM}}{d M_{\rm DM}}\right\vert_{\rm Fig13}
$$
is the WD number density distribution near the solar system, and $dN_{\rm WD}/dM_{\rm WD}|_{\rm Fig13}$ 
is the WD number distribution adopted from Fig.13 of Ref.~\cite{Kilic_2020} within $100~{\rm pc}$ near the solar system.
The total event rate $N_\text{event rate}$ of SN Ia explosion induced by transiting PBHs can be obtained after integrating the Milky Way volume and WD mass distribution:
\begin{eqnarray}
&& N_\text{event rate}(M_{\rm PBH},f_{\rm PBH}) \nonumber \\
&&~~~~=
\int^{30~{\rm kpc}}_0 2\pi R_{\rm MW} dR_{\rm MW} 
\int^{1~{\rm kpc}}_{-1~{\rm kpc}} dZ_{\rm MW} 
\int^{1.40 M_\odot}_{0.85 M_\odot} dM_{\rm WD}\, \nonumber \\
&&~~~~~~~~~~~
\frac{dn_{\rm WD}}{dM_{\rm WD}}(R_{\rm MW},Z_{\rm MW})
\times 
\Gamma(\rho_{\rm DM}(r_{\rm MW}),M_{\rm WD},M_{\rm PBH},f_{\rm PBH}) \,. \nonumber 
\end{eqnarray}
The integration of $M_{\rm WD}$ starts from $0.85\,M_{\odot}$ because Type Ia supernovae can be observed only if $M_{\rm WD}$ is greater than $0.85\,M_{\odot}$~\cite{Graham:2015apa}.
For certain sub-Chandrasekhar WD mass, the SN Ia explosion rate induced by transiting PBHs is given by~\cite{Steigerwald:2021vgi}
\begin{eqnarray}
\label{eq:capital gamma}
\Gamma\left(\rho_{\rm DM},v_{*\rm esc}(M_{\rm WD},R_m),R_m(M_{\rm WD},M_{\rm PBH}),M_{\rm PBH},f_{\rm PBH} \right) = \frac{\pi R_m^2 v_{*\rm esc}^2  }{v_{\rm gal}}
\left( \frac{\rho_{\rm DM} f_{\rm PBH}}{M_{\rm PBH}} \right)\,, \nonumber \\
\end{eqnarray}
where $R_m$ indicates the effective radius to trigger the detonation ignition, when a PBH transits through a WD~\cite{Steigerwald:2021vgi}. 
Non-zero $R_m$ requires the $ 3\times 10^{-13} M_\odot \lesssim M_{\rm PBH} \lesssim 5\times 10^{-9} M_\odot $ and 
$0.7 M_\odot \lesssim M_{\rm WD} \lesssim 1.4 M_\odot $.
The mean velocity $v_{\rm gal}$ of PBHs under thermal motion in the galaxy is $2\times10^5~{\rm m/s}$, and the WD escape velocity at $R_m$, $v_{*{\rm esc}}$, is computed by taking into account the WD mass-dependent density profiles, shown in Appendix~\ref{sec:appendix}. Fig.~\ref{fig:SN_event_rate} presents the SN Ia event rate derived from Eq.(\ref{eq:DM density profile}) to Eq.(\ref{eq:capital gamma}) in the NFW, Isothermal, Kra and Moore profiles for $f_{\rm PBH}$ equals to unity. We see the event rates highly depend on the PBH density near galaxy center, and vary within in an order of magnitude associating with different DM profiles. Comparing left and right panels, the uncertainty from detonation criteria $\bar{\alpha}$, Eq.(\ref{eq:alpha}), can also change the event rate by one order. The grey band presents the observed SN Ia Milky Way event rate which is $0.54\pm0.12$ per century with 68.3\% confidence~\cite{deJong:2009iq}. 
In NFW profile with $\bar{\alpha}=1$ ($\bar{\alpha}=10$) case, $M_{\rm PBH}\approx 2\times 10^{-12}M_\odot$ ($M_{\rm PBH}\approx 4\times 10^{-12}M_\odot$) reproduces the observed event rate.
%

\begin{figure}[t!]
\centering
\includegraphics[height=2.2in,angle=0]{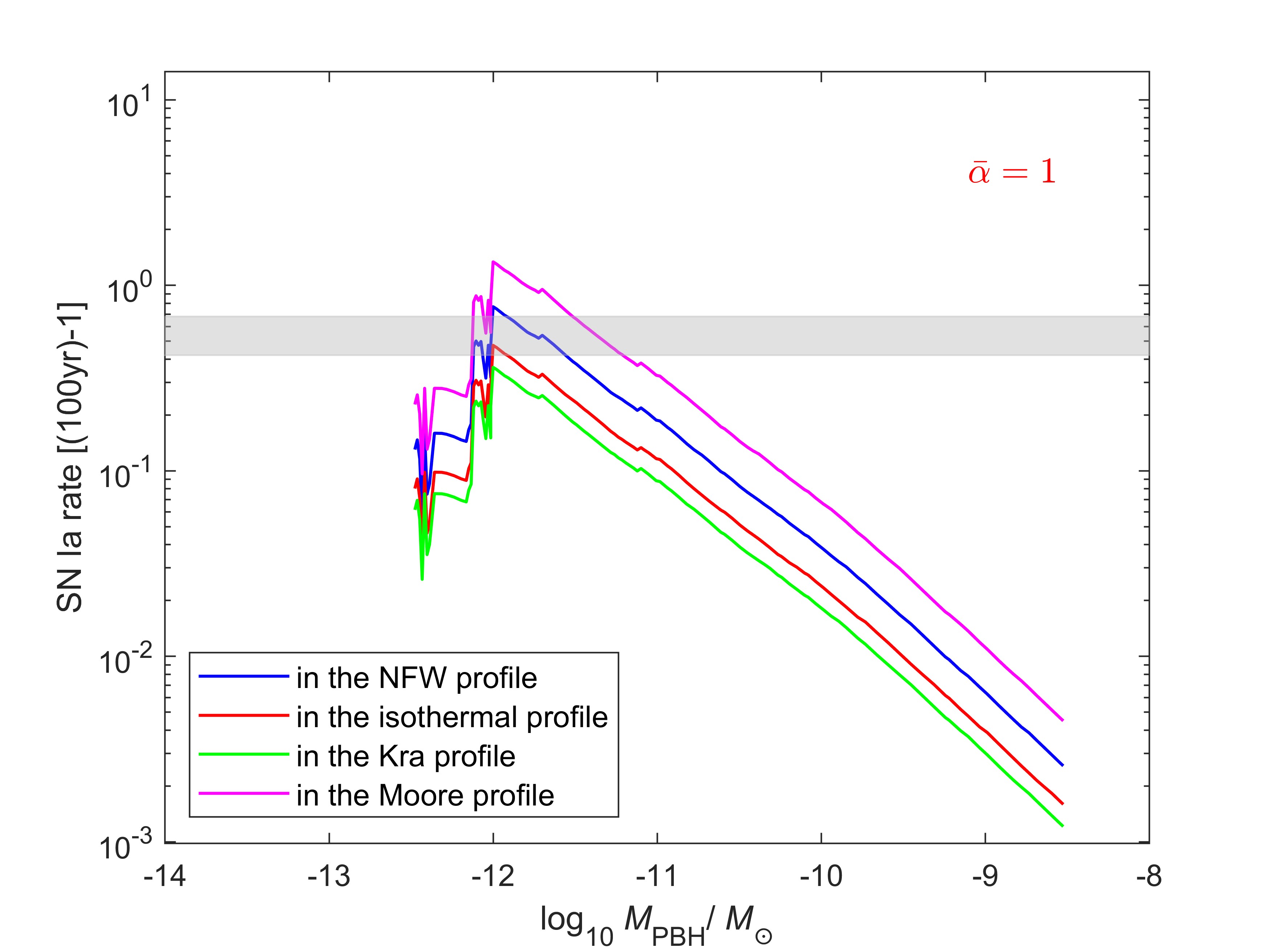}
\includegraphics[height=2.2in,angle=0]{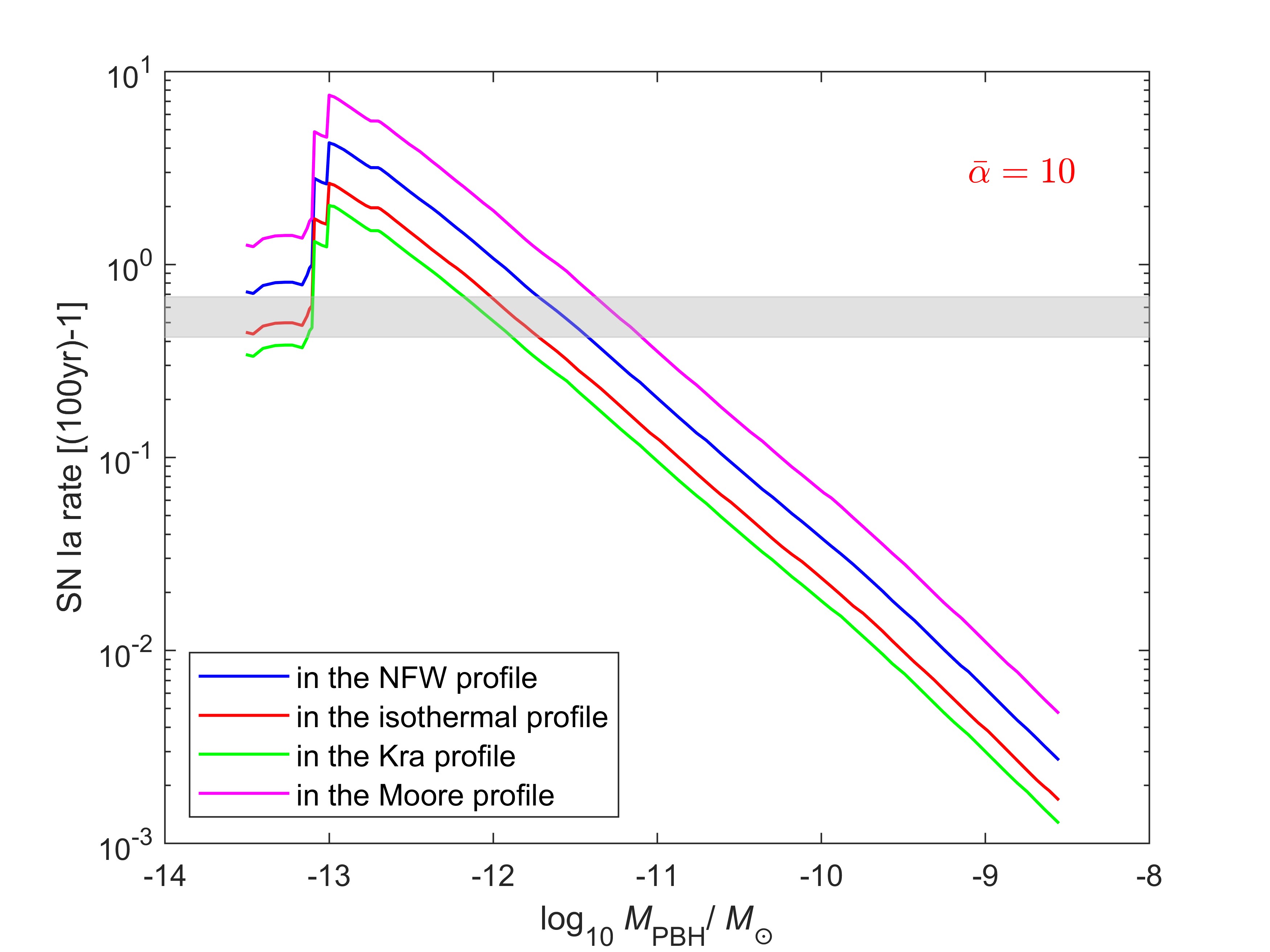}
\caption{\small \label{fig:SN_event_rate}
SN Ia event rate induced by PBH in the NFW, isothermal, Kra and Moore profiles when $f_{\rm PBH}$ is 1 and $\bar{\alpha}$ is 1 (left) and 10 (right).
}
\end{figure}

\bigskip

\section{Constraints from SN Ia observations}
\label{sec:even_rate}

\begin{figure}[t!]
\centering
\includegraphics[height=2.2in,angle=0]{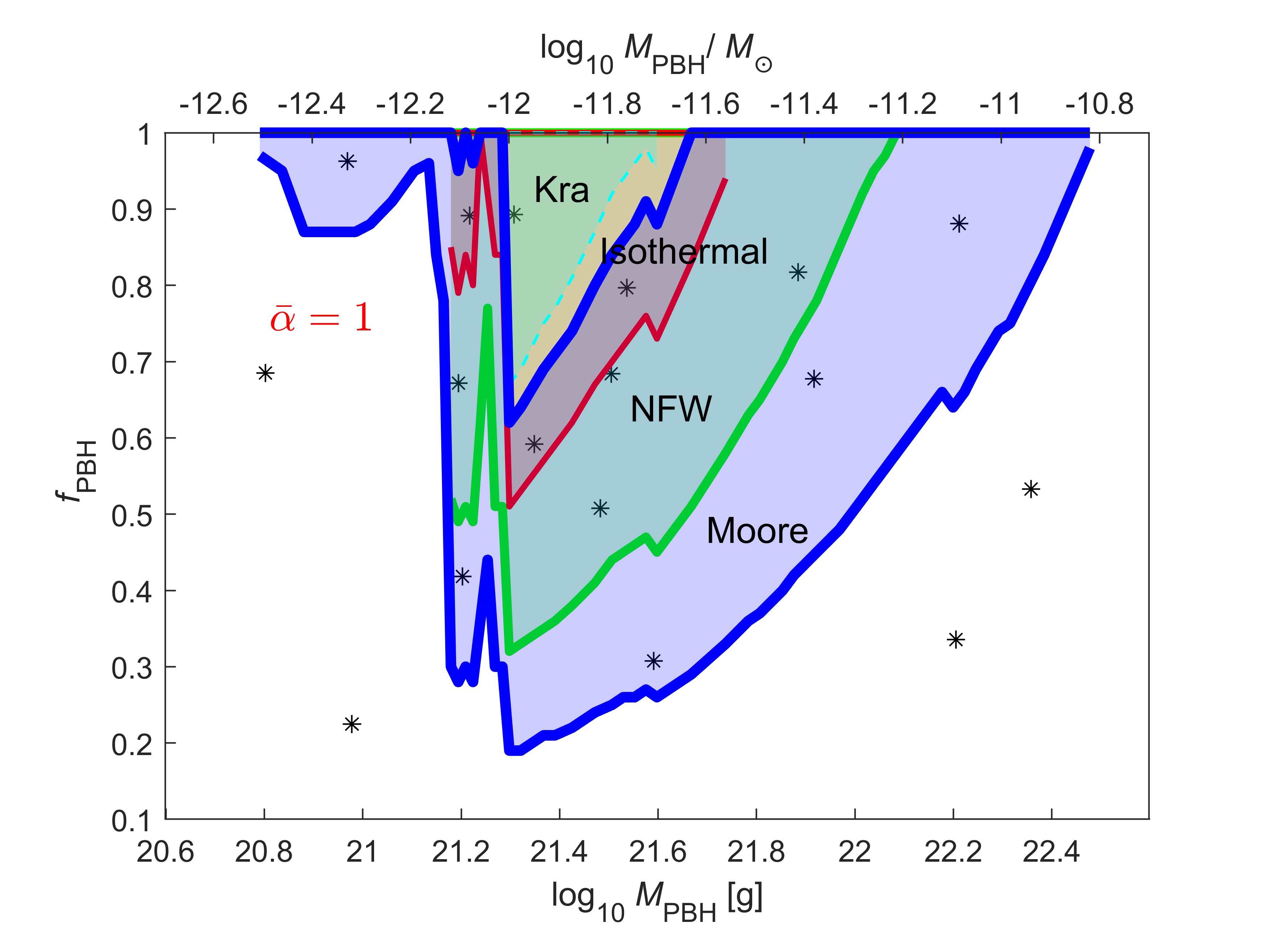}
\includegraphics[height=2.2in,angle=0]{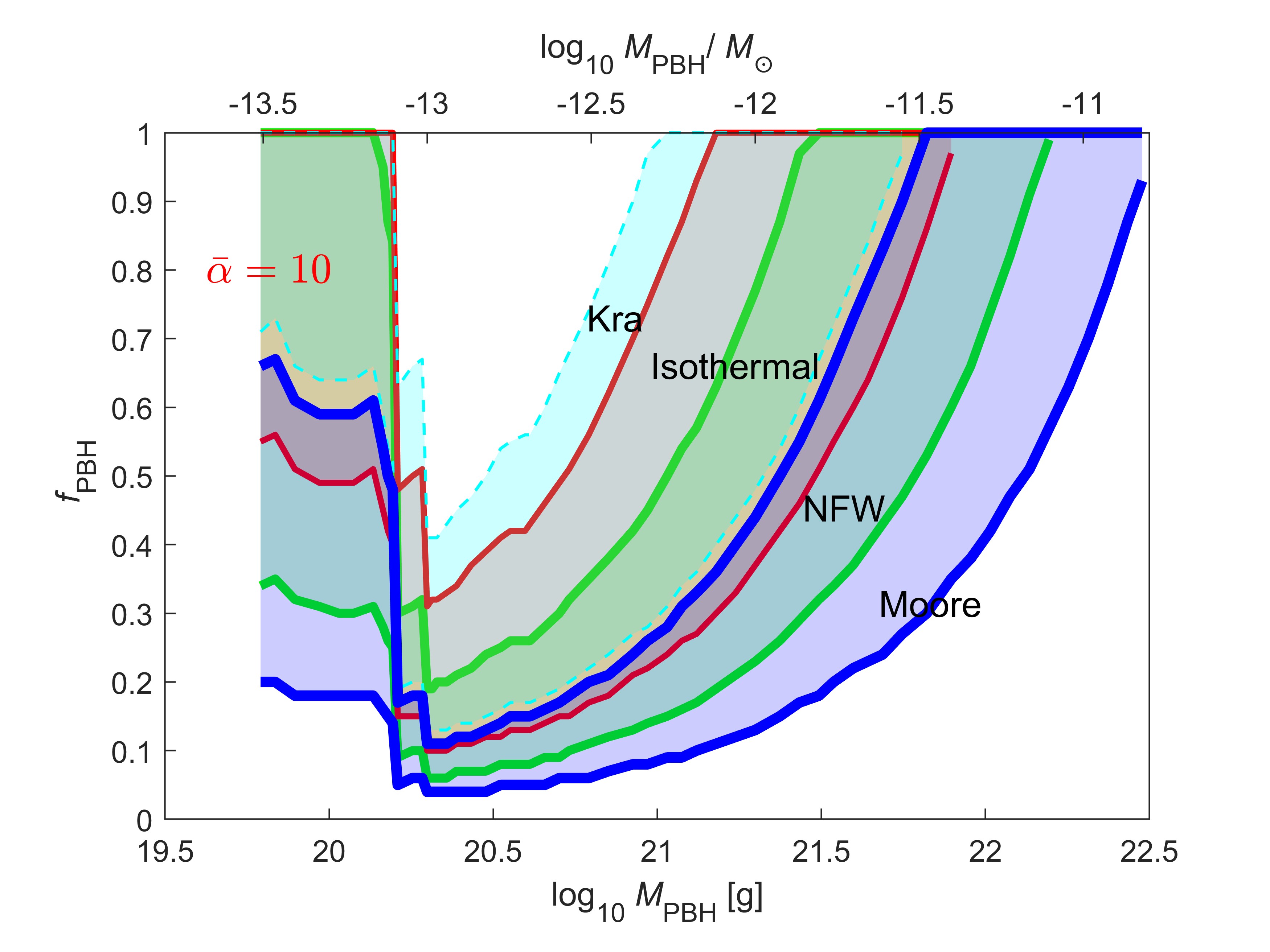}
\caption{\small \label{fig:f_PBH}
The shaded parameter regions reproduce the observed SN Ia event rate in the NFW (green), isothermal (red), Kra (cyan) and Moore (blue) profiles, respectively, when $\bar{\alpha}$ is 1 (left) and 10 (right).
The "*" points within the shaded regions indicate the {\bf BP}s belong to particular profile in Table~\ref{table:1} and ~\ref{table:2}, otherwise, they are {\rm BP}-1 to 4 in Table~\ref{table:2}.
}
\end{figure}

Varying the PBH abundance, the shaded regions in Fig.~\ref{fig:f_PBH} show the preferred parameter space on PBHs from observed SN Ia event rate
$0.54\pm 0.12$ per century, with two-dimensional 95\% confidence level on $(M_{\rm PBH},f_{\rm PBH})$ panel. 
In particular the relevant parameter space, 
$M_{\rm PBH}$ falls in the interval $[3.1\times 10^{-13}\,M_{\odot},1.5\times 10^{-11}\,M_{\odot}]$ for $\bar{\alpha} = 1$ and $[3.1\times 10^{-14}\,M_{\odot},1.5\times 10^{-11}\,M_{\odot}]$ for $\bar{\alpha} = 10$ with $f_{\rm PBH}$ in between [0.05, 1]. The corresponding values of $f_{\rm PBH}$ above these shaded regions exceedingly produce the SN Ia event rate and thereby are excluded, therefore only $10^{-12}\leq M_{\rm  PBH}/M_\odot \leq 2.5\times 10^{-12}$ with $f_{\rm PBH}=1$  and $\bar{\alpha}=1$ under the Moore profile is disfavored.
However for $\bar{\alpha}=10$, 
$10^{-13}\leq M_{\rm  PBH}/M_\odot \leq 4\times 10^{-13}$ with $f_{\rm PBH}=1$ is excluded under all profiles, which is currently insensitive by neither PBH evaporations nor microlensing observations.

For $\bar\alpha = 1$, when $M_{\rm PBH}$ is greater than $1.5\times 10^{-11}\,M_{\odot}$, there is no SN Ia event: because the number density $n_{\rm PBH}$ of PBHs decreases as $M_{\rm PBH}$ increases, while keeping $\rho_{\rm DM}$ intact, the PBH-induced SN Ia rate decreases.
On the other hand, the required minimum PBH mass to trigger the detonation ignition are $3.1\times 10^{-13} M_\odot$ ($3.1\times 10^{-12} M_\odot$) for $\bar{\alpha}=1$ ($\bar{\alpha}=10$)~\cite{Steigerwald:2021vgi}, due to the fact that the narrowing regime of $M_{\rm WD}$ corresponding to non-zero $R_m$.
As a result, the shaded regions in Figure \ref{fig:f_PBH} terminates at the left-handed side when $M_{\rm PBH}$ is less than aforementioned minimal values.

There are several caveats of estimating the SN Ia event rate. We adopted the observed WD mass distribution within 100 pc, then extrapolated to whole galaxy via assuming correlated mass distributions of WD and stellar. Furthermore, we only considered PBH intercepts via isolate WD, however, about 10\% to 30\% WD population constitute in binary system~\cite{WD_binary1,WD_binary2}, which may alter the event rate.
Three criteria of thermal nuclear detonation are base on the carbon-oxygen WD core assumption. Even though this percentage is higher for heavier WD mass, but there is unknown fraction of oxygen-neon-magnesium core of WD heavier than 1.05$M_\odot$. Among three criteria, the ratio between the critical induction length and impact parameter is an order one quantity, for which, we demonstrated the uncertainties by setting $\bar{\alpha}=1$ and $\bar{\alpha}=10$.
The last not the least, the event rate highly depends on the PBH distribution profile, and four different profiles: NFW, isothermal, Kra, and Moore, are used to investigate the uncertainties.

   
\begin{figure}[t!]
\centering
\includegraphics[height=2.9in,angle=-90]{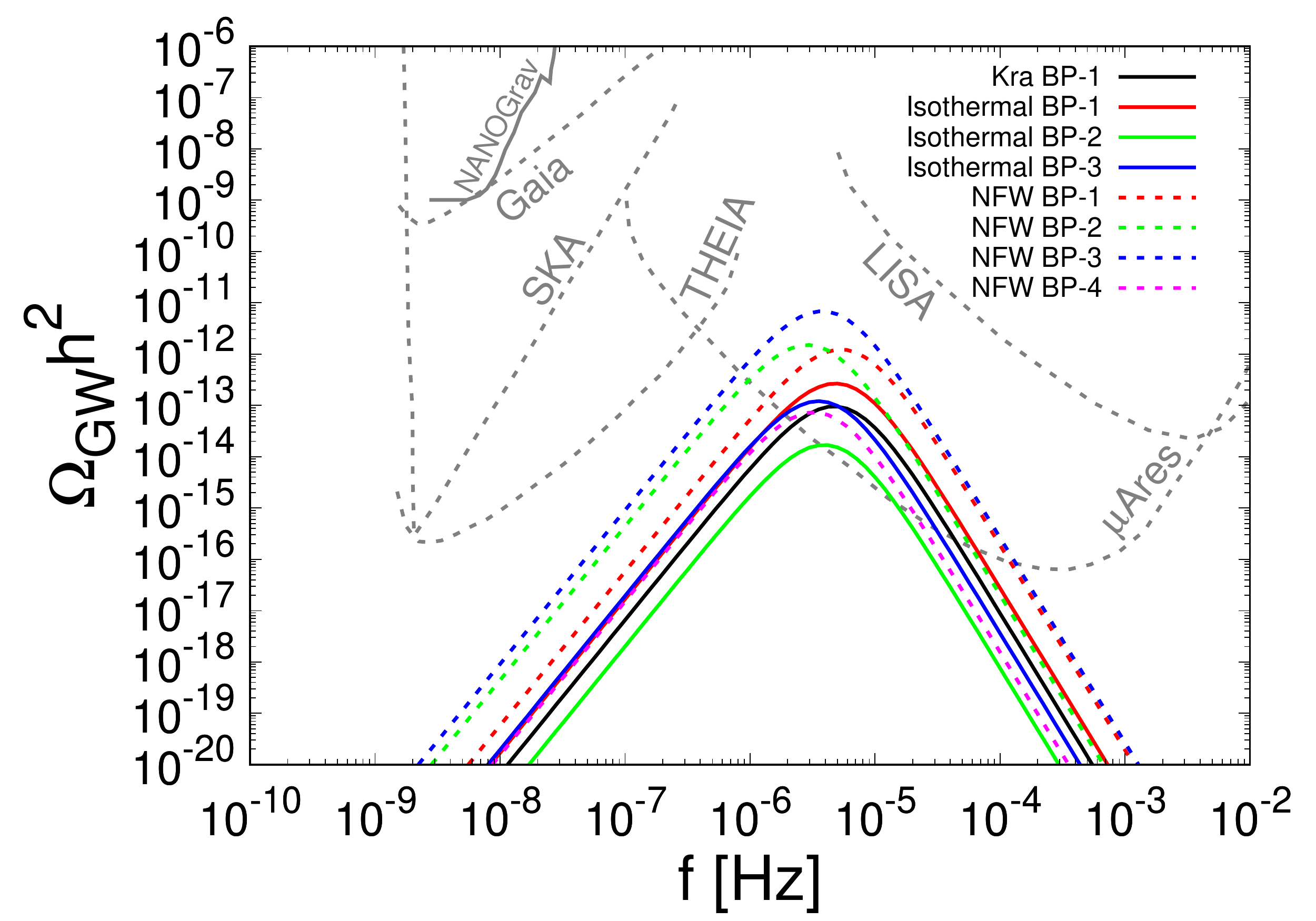}
\includegraphics[height=2.9in,angle=-90]{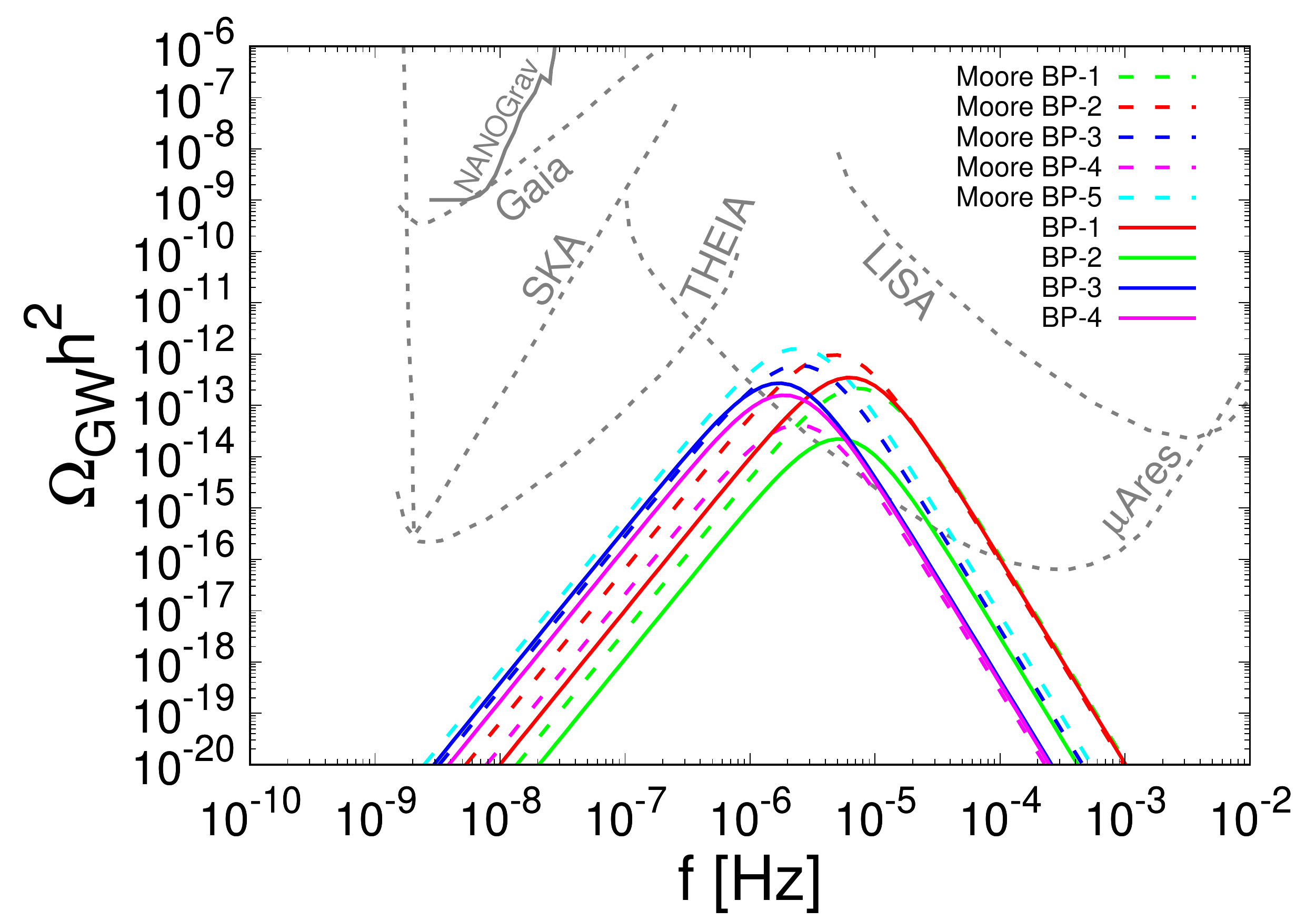}
\caption{\small \label{fig:WD_f_Oh2_2}
Gravitational wave power spectra for the benchmark points in Table~\ref{table:1} (left) and Table~\ref{table:2} (right).
}
\end{figure}

Table~\ref{table:1} and Table~\ref{table:2} illustrate the parameters of the selected benchmark points ({\bf BP}s), and reveal the common features in the following. The frequency of the gravitational waves for the {\bf BP}s is in the order $10^{-6}$ to $10^{-5}$ Hz, since it is proportional to $\beta$, the duration of first order phase transition happened at $T_\star$, and then red shifted to the present epoch. 
The correlated GW spectra of {\bf BP}s are present in Figure \ref{fig:WD_f_Oh2_2}.
There is not much deviation among the BPs GW spectra, where the peak frequencies are in between $10^{-6}$ to $10^{-5}$ Hz, which will only be sensitive by $\mu$Ares~\cite{Sesana:2019vho}.
On the other hand, the energy scale of FOPT, $B^{1/4}$, is in the order MeV, which corresponds to $M_{\rm PBH}/M_{\odot}$ in the range from $\mathcal{O}(10^{-13})$ to $\mathcal{O}(10^{-12})$, in which the mass is too heavy to have substantial Hawking evaporation, meanwhile too small to produce microlensing signal. Regard the stability of FB, it is found that the bare mass of $\chi$ in the false vacuum, $m$, should be greater than zero , and the $g_\chi\simeq \mathcal{O}(1)$, otherwise the stability condition for the fermi balls, first inequality in Eq.(\ref{eq:stability}),
will not be satisfied. Furthermore, the asymmetry $\eta_{\rm DM}\sim \mathcal{O}(10^{-8})$ suppress PBH relic abundance thus avoid PBH overclosing the universe.
Additional constraint comes from the effective number of extra neutrino species contributed by the dark sector, $\Delta N_{\rm eff}$. Combining the observations of cosmic microwave background, baryon acoustic oscillations and Big
Bang Nucleosynthesis give the upper limit $\Delta N_{\rm eff}\leq 0.55$~\cite{Planck:2018vyg,Mangano:2011ar}. But since the energy scale of FOPT of {\bf BP}s are all above MeV,
we conservatively require the $\Delta N_{\rm eff}\leq 0.5$ by considering thermally decoupled SM and dark sectors with temperature ratio $r_T=T_{\rm dark}/T_{\rm SM}$, where the lower $T_{\rm dark}$ suppresses the $\Delta N_{\rm eff}$.

    \begin{table}[h]
        \centering
        \resizebox{\textwidth}{!}{
        \begin{tabular}{c|c|c c c|c c c c}
            \hline
            \hline
            & \bf{Kra BP-1} & \bf{Isothermal BP-1} & \bf{Isothermal BP-2} & \bf{Isothermal BP-3}& \bf{NFW BP-1} & \bf{NFW BP-2} & \bf{NFW BP-3} & \bf{NFW BP-4} \\
           \hline
           \hline
            $B^{1/4}/\rm MeV$ & $4.316$ & $5.477$ & $3.262$ & $2.973$ & $11.72$ & $5.269$ & $9.840$ & $4.433$ \\
            
            $\lambda$ & $0.120$ & $0.104$ & $0.089$ & $0.059$ & $0.057$ & $0.186$ & $0.086$ & $0.123$ \\
            
            $D$ & $0.794$ & $1.243$ & $0.922$ & $1.674$ & $0.629$ & $1.891$ & $1.564$ & $0.434$ \\
            
            $\eta_{\mathrm{DM}}$ & $2.97\times10^{-8}$ & $2.73\times10^{-8}$ & $2.78\times10^{-8}$ & $3.66\times10^{-8}$ & $5.89\times10^{-9}$ & $1.54\times10^{-8}$ & $9.36\times10^{-9}$ & $1.87\times10^{-8}$ \\
            
            $r_T$ & $0.506$ & $0.434$ & $0.320$ & $0.379$ & $0.353$ & $0.433$ & $0.361$ & $0.398$ \\
            
            $C/\rm MeV$ & $0.203$ & $0.240$ & $0.110$ & $0.032$ & $0.343$ & $0.556$ & $0.550$ & $0.353$ \\
            
            $g_{\chi}$ & $1.520$ & $1.329$ & $1.274$ & $0.807$ & $0.833$ & $1.352$ & $0.822$ & $1.459$ \\
            
            $m/B^{1/4}$ & $0.059$ & $0.020$ & $0.383$ & $1.508$ & $0.936$ & $0.093$ & $0.451$ & $0.699$ \\
            \hline
            $M_\mathrm{PBH}/M_{\bigodot}$ & $1.02\times10^{-12}$ & $8.29\times10^{-13}$ & $1.12\times10^{-12}$ & $1.73\times10^{-12}$ &  $7.87\times10^{-13}$ & 
            $1.53\times10^{-12}$ &
            $1.61\times10^{-12}$ &
            $3.85\times10^{-12}$ \\
            
            $f_\mathrm{PBH}$ & $0.893$ & $0.891$ & $0.592$ & $0.797$ & $0.671$ & $0.508$ & $0.684$ & $0.817$ \\

            $R(T_\star)\, [\rm GeV^{-1}]$ & $7.23\times10^{19}$ & $6.13\times10^{19}$ & $8.18\times10^{19}$ & $1.61\times10^{20}$ & $1.84\times10^{19}$ & $1.15\times10^{20}$ & $5.00\times10^{19}$ & $6.43\times10^{19}$\\

            $vw$ & $0.880$ & $0.943$ & $0.930$ & $0.955$ & $0.977$ & $0.969$ & $0.990$ & $0.902$\\

            $R_{\rm FB}\, [\rm GeV^{-1}]$ & $9.29\times10^{17}$ &	$7.35\times10^{17}$ &	$1.37\times10^{18}$ & $1.73\times10^{18}$ & $1.29\times10^{17}$ & $9.83\times10^{17}$ &	$3.28\times10^{17}$ &	$8.23\times10^{17}$\\

            $Q_{\rm FB}$ & $9.50\times10^{46}$ & $7.05\times10^{46}$ & $1.40\times10^{47}$ &	$2.17\times10^{47}$ & $1.89\times10^{46}$ &	$1.32\times10^{47}$ & $6.37\times10^{46}$ & $2.38\times10^{47}$\\
            
          \hline
        \end{tabular}
        }
        \caption{
        The {\bf BP}s in the Kra, isothermal and NFW profiles, which form PBHs after FOPT with $A=0.1$ fixed.
        }
        \label{table:1}
    \end{table}

    \begin{table}[h]
        \centering
        \resizebox{\textwidth}{!}{
        \begin{tabular}{c|c c c c c|c c c c}
            \hline
            \hline
            & \bf{Moore BP-1} & \bf{Moore BP-2} & \bf{Moore BP-3} & \bf{Moore BP-4}& \bf{Moore BP-5} & \bf{BP-1} & \bf{BP-2} & \bf{BP-3} & \bf{BP-4} \\
           \hline
           \hline
            $B^{1/4}/\rm MeV$ & $13.27$ & $8.947$ & $3.605$ & $1.214$ & $4.809$ & $9.572$ & $7.175$ & $2.030$ & $3.726$\\
            
            $\lambda$ & $0.166$ & $0.056$ & $0.184$ & $0.135$ & $0.085$ & $0.153$ & $0.167$ & $0.072$ & $0.199$\\
            
            $D$ & $0.282$ & $0.430$ & $3.088$ & $3.779$ & $0.991$ & $1.025$ & $0.195$ & $0.756$ & $0.708$ \\
            
            $\eta_{\mathrm{DM}}$ & $4.93\times10^{-9}$ & $4.04\times10^{-9}$ & $1.65\times10^{-8}$ & $1.48\times10^{-7}$ & $2.43\times10^{-8}$ & $8.37\times10^{-9}$ & $1.94\times10^{-9}$ &  $2.20\times10^{-8}$ & $1.87\times10^{-8}$ \\
            
            $r_T$ & $0.551$ & $0.421$ & $0.329$ & $0.440$ & $0.367$  & $0.419$ & $0.441$ & $0.421$ & $0.334$\\
            
            $C/\rm MeV$ & $1.747$ & $0.159$ & $0.357$ & $0.043$ & $0.247$ & $0.864$ & $1.114$ & $0.051$ & $0.587$\\
            
            $g_{\chi}$ & $1.190$ & $0.923$ & $1.445$ & $1.282$ & $1.129$ & $1.156$ & $1.514$ & $1.169$ & $1.837$\\
            
            $m/B^{1/4}$ & $2.265$ & $1.168$ & $0.027$ & $0.020$ & $0.590$ & $1.400$ & $1.105$ & $0.375$ & $0.012$ \\
            \hline
            $M_\mathrm{PBH}/M_{\bigodot}$ & $4.68\times10^{-13}$ & $8.02\times10^{-13}$ & $1.96\times10^{-12}$ & $4.15\times10^{-12}$ & $8.19\times10^{-12}$ &  $3.19\times10^{-13}$ & $4.78\times10^{-13}$ & $8.06\times10^{-12}$ & $1.15\times10^{-11}$ \\
            
            $f_\mathrm{PBH}$ & $0.963$ & $0.418$ & $0.308$ & $0.677$ & $0.881$ & $0.685$ & $0.225$ & $0.335$ & $0.533$\\
            
            $R(T_\star)\, [\rm GeV^{-1}]$ & $9.76\times10^{18}$ & $2.51\times10^{19}$ & $2.11\times10^{20}$ & $8.60\times10^{20}$ & $1.23\times10^{20}$ & $2.41\times10^{19}$ & $1.71\times10^{19}$ & $3.69\times10^{20}$ & $1.50\times10^{20}$\\

            $vw$ & $0.850$ & $0.952$ & $0.984$ & $0.925$ & $0.977$ & $0.945$ & $0.818$ & $0.941$ & $0.952$\\

            $R_{\rm FB}\, [\rm GeV^{-1}]$ & $5.87\times10^{16}$ & $1.52\times10^{17}$ & $2.06\times10^{18}$ & $1.89\times10^{19}$ & $1.31\times10^{18}$ & $1.70\times10^{17}$ & $1.14\times10^{17}$ & $4.21\times10^{18}$ & $1.95\times10^{18}$\\

            $Q_{\rm FB}$ & $6.71\times10^{45}$ & $2.14\times10^{46}$ & $2.90\times10^{47}$ & $2.50\times10^{48}$ & $6.27\times10^{47}$ & $1.07\times10^{46}$ & $1.12\times10^{46}$ & $1.45\times10^{48}$ & $1.08\times10^{48}$ \\
            
          \hline
        \end{tabular}
        }
        \caption{
        The {\bf BP}s in the Moore profile and outside of all the profiles, which form PBHs after FOPT with $A=0.1$ fixed.
        }
        \label{table:2}
    \end{table}



\bigskip

\section{Summary}
\label{sec:summary}

We re-examine the SN Ia event rate triggered by PBHs transiting through sub-Chandrasekhar WDs. In particular, we extrapolated the WD distribution to the whole Milky Way via adopting the observation data of WD mass distribution within 100 pc and assumed the WD number density is proportional to stellar one. 
From our calculation, for $f_{\rm PBH}\leq 1$,
the SN Ia observed event rate $0.54\pm 0.12$ per century, prefers PBH mass regions in $[7.6\times10^{-13}\,M_{\odot},6.1\times10^{-12}\,M_{\odot}]$,
$[7.6\times10^{-13}\,M_{\odot},2.7\times10^{-12}\,M_{\odot}]$, 
$[1.0\times10^{-12}\,M_{\odot},2.0\times10^{-12}\,M_{\odot}]$ and
$[3.1\times10^{-13}\,M_{\odot},1.5\times10^{-11}\,M_{\odot}]$
respectively under the assumptions of NFW, isothermal, Kra and Moore profiles.
For $\bar{\alpha}=1$, only under the Moore profile, $10^{-12} \leq M_{\rm PBH}/M_\odot \leq 2.5\times 10^{-12}$ with $f_{\rm PBH}=1$ predicts exceeding event rate and thus is disfavored.
However for $\bar{\alpha}=10$, 
$10^{-13}\leq M_{\rm  PBH}/M_\odot \leq 4\times 10^{-13}$ with $f_{\rm PBH}=1$ is excluded under all profiles, and this PBH mass window is currently insensitive by neither PBH evaporation nor microlensing observations.
Next, we consider the PBH production mechanism from cosmological FOPT in the dark sector to realize the required mass and abundance, which corresponds to the energy difference between false and true vacuum at zero temperature to be $1 \lesssim B^{1/4}/{\rm MeV}\lesssim 10$. 
In addition, the signal of gravitational wave with peak frequencies between $10^{-6}$ and $10^{-5}$ Hz from FOPT can be detected by future $\mu$Ares telescope.

\bigskip
\section*{Acknowledgements}  
PYT thank Prof. Hsiang-Kuang Chang for useful discussions of white dwarf and Type Ia supernova. PYT is supported in part by NSTC (National Science and Technology Council) Grant No. (111-2112-M-007-012-MY3). PJC is supported by Ministry of Education with Grant No. 110J0352I4.

\bigskip

\appendix
\section{White dwarf density profile and escape velocity}
\label{sec:appendix}

We derive the white dwarf density profile according to Ref.~\cite{WD_profile} by solving the coupled differential equations
\begin{eqnarray}
\label{eq:WD}
&& \frac{dm}{dr}=4\pi r^2 \rho\,, \nonumber \\
&& \frac{d\rho}{dr}=-\left( \frac{dP}{d\rho} \right)^{-1} \frac{G_N m}{r^2} \rho\,,
\end{eqnarray}
where $m$ is the mass of the portion of WD from its centre to the shell with radius $r$, $\rho$ and $P$ are the density and pressure of the WD, respectively, and $G_{N}$ is the gravitational constant. The equation of state in WD can be approximated as a relativistic free Fermi gas as following
\begin{eqnarray}
\label{eq:eq of state}
\frac{dP}{d\rho}&=& Y_e \frac{m_e c^2}{m_p} \gamma\left( \frac{\rho}{\rho_0} \right)\,, \nonumber \\
\rho_0 & \equiv & \frac{m_p m^3_e}{3 \pi^2 Y_e}
\end{eqnarray}
with $Y_e$ the number of electrons per nucleon (for $^{12}C$, $Y_e=1/2$), $m_{e}$ and $m_{p}$ the electron and proton mass, respectively, $\rho_0$ the density at center (as initial condition eventually determine the total WD mass OR as the initial condition is determined by the total WD mass $M_{\rm WD}$),
and $\gamma$ function
\begin{eqnarray}
&& \gamma(y)=\frac{y^{2/3}}{3(1+y^{2/3})^{1/2}}\,, \nonumber
\end{eqnarray}

\begin{figure}[t!]
\centering
\includegraphics[height=2.5in,angle=0]{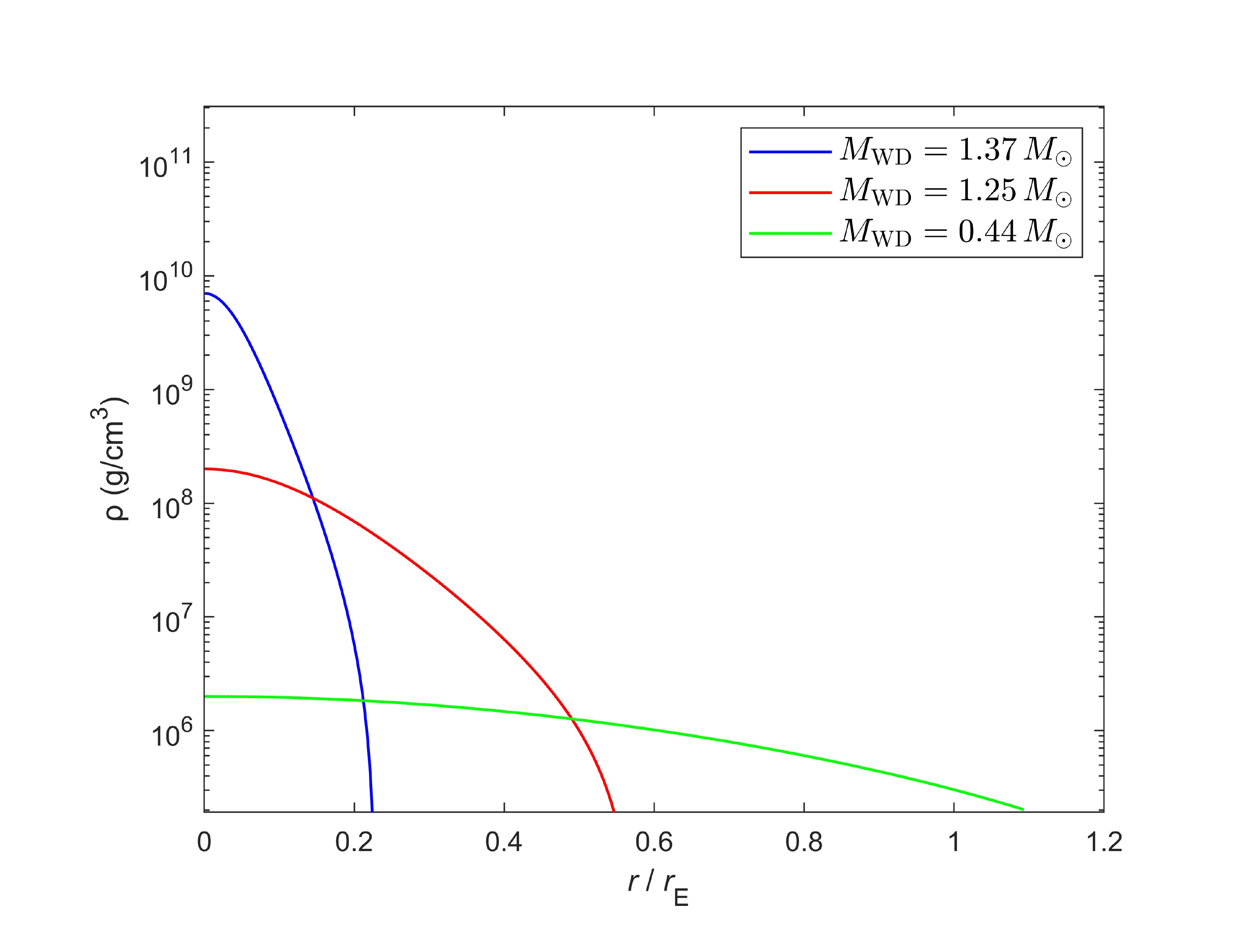}
\caption{\small \label{fig:WD_profile}
WD density profiles of three different WD masses, where $r_E$ is Earth's radius.
}
\end{figure}

Eq.(\ref{eq:WD}) can be numerically solved by the Runge-Kutta method,
then we can obtain WD density profile, $\rho(r)$, 
with respect to different WD masses, $M_{\rm WD}$.
Figure \ref{fig:WD_profile} shows the WD density profiles of three different WD masses obtained from Eq.(\ref{eq:WD}) and Eq.(\ref{eq:eq of state}), which can be compared with results from Fig.1 of Ref.~\cite{Biermann:2010xm}.

The gravitational potential of WD produces focusing effect, which is able to increase the effective cross section with incoming PBHs flux. The cross section enhancement is proportional to the square of ratio of escape velocity to the mean galactic velocity.
For $r \geq R_{\rm WD}$, where $R_{\rm WD}$ is the radius of the WD, the escape velocity at $r$ from the WD is given by
$$
v_{*\rm esc}(r)=\left( \frac{2 G_N M_{\rm WD}}{r} \right)^{1/2}\,.
$$
For $r < R_{\rm WD}$, the escape velocity at $r$ from the WD is given by
\begin{eqnarray}
&& v_{*\rm esc}(r)=\left( \frac{2G_N M_{\rm WD}}{R_{\rm WD}}
+\int^{R_{\rm WD}}_r \frac{2G_N m(r')}{r'^2}dr'  \right)^{1/2} \,, \\
&& \bar{v}_{*\rm esc} \equiv  v_{*\rm esc}|_{r=0}\,. \nonumber 
\end{eqnarray}

\bigskip

\bigskip

\end{document}